\documentclass[]{aa} 
\usepackage{graphicx}
\usepackage{txfonts}
\usepackage[T1]{fontenc}
\usepackage{pdflscape}
\usepackage{rotating}
\usepackage{float}
\usepackage{graphicx}
\usepackage{caption}
\usepackage{subcaption}
\usepackage{longtable} 
\usepackage{booktabs} 
\usepackage{pdfpages}
\usepackage{pgffor}
\usepackage{hyperref}

\begin{document} 
\title{Evolutionary Map of the Universe (EMU): the fraction of radio-detected quasi-stellar objects at high redshift}
\titlerunning{The fraction of radio-detected QSOs at high-z}
   \author{A. Caccianiga\fnmsep\thanks{alessandro.caccianiga@inaf.it}
          \inst{1}
          \and
          D. Dallacasa
          \inst{2,3}
          \and
          L. Ighina
          \inst{1,4}
          \and
          A. Moretti
          \inst{1}
          \and
          J. Afonso
          \inst{5,6}
          \and
          S. Ant\'on
          \inst{7}
          \and
          R. Carvajal
          \inst{5}
          \and
          Y. Gordon
           \inst{8}
          \and
          A. Hopkins,
           \inst{9}
          \and
          M.J. March\~a
          \inst{10}
          \and
          R. Norris
          \inst{11}
          \and
          F. Rahman
          \inst{12}
          \and
          G. F. Ramos
          \inst{7}
          \and
          N. Seymour
          \inst{13}
          \and
          C. Spingola
          \inst{3}
          \and
          J. Th. van Loon
          \inst{14}
          }

   \institute{
   INAF - Osservatorio Astronomico di Brera, via Brera 28,  I-20121 Milan, Italy
              \email{alessandro.caccianiga@inaf.it}
         \and
         Dipartimento di Astronomia, Universit\`a di Bologna, via Ranzani 1, 40127, Bologna, Italy
         \and
         INAF - Istituto di Radioastronomia, Via Gobetti 101, I-40129 Bologna, Italy
         \and
         Center for Astrophysics—Harvard \& Smithsonian, 60 Garden Street, Cambridge, MA 02138, USA
         \and
          Instituto de Astrof\'isica e Ci\^encias do Espaço, Universidade de Lisboa, OAL, Tapada da Ajuda, PT-1349-018 Lisbon, Portugal
          \and
         Departamento de F\'isica, Faculdade de Ci\^encias, Universidade de Lisboa, Edif\'icio C8, Campo Grande, PT-1749-016 Lisbon, Portugal
         \and
         Centro de F\'isica da UC, Departamento de Física, Universidade de Coimbra, 3004-516 Coimbra, Portugal
        \and
        Department of Physics, University of Wisconsin-Whitewater, 800 W. Main St., Whitewater, WI 53190, USA.
        \and
        School of Mathematical and Physical Sciences, 12 Wally’s Walk, Macquarie University, NSW 2109, Australia
         \and
         Department of Physics \& Astronomy, University College London, Gower Street, London WC1E 6BT, UK
         \and
         CSIRO Space \& Astronomy, PO Box 76, Epping, NSW 1710, Australia
         \and
        Syed Babar Ali School of Science and Engineering (SBASSE), Lahore University of Management Sciences (LUMS), Lahore, Pakistan
         \and
        International Centre for Radio Astronomy Research, Curtin University, GPO Box U1987, Bentley, WA 6845, Australia
          \and
          Lennard-Jones Laboratories, Keele University, ST5 5BG, UK
             }

   \date{Received XX; accepted yy}

 
  \abstract
   {Determining how frequently relativistic jets occurred in the young Universe is essential to assess their impact on the early evolution of supermassive black holes and their host galaxies.
   }
   {The aim of this paper is to use the Evolutionary Map of the Universe (EMU), which provides a deep radio look at $\sim$1~GHz on a large sky area (about 9000 deg$^2$ at the end of February 2026), to assess the fraction of high-z quasi-stellar objects (QSOs) that are radio-emitting at $\sim$0.2~mJy/beam level and to establish if the radio emission is mainly produced by relativistic jets.}
   {A sample of 808 type~1 QSOs in the redshift range 4$<$z$<$9 and covering the sky area currently observed by EMU was compiled from the literature and the SDSS spectroscopic database and cross-matched with EMU source catalogue.
   In total, 124 high-z QSOs are detected by EMU, corresponding to 15\% of the sample. The large majority (70\%) of the QSOs radio detected by EMU are reported in the radio for the first time. In particular, we report 13 new radio detections at 5.5$\leq$z$\leq$6.5. We use this sample to establish the detection rate at different optical magnitudes. We also study the radio luminosity of these QSOs and their radio-loudness (R), defined as the radio-to-optical luminosity ratio at 5~GHz and 4400\AA\  respectively, to infer the possible origin of the observed radio emission.}
   {We show that the fraction of radio-loud (R$>$10) high-z QSOs is about 8\% while the fraction of radio detections is a strong function of the optical magnitude and reaches values of $\sim$35\% when only the optically brightest (mag$\sim$18) QSOs are considered.
   Since all the radio detected QSOs, including those classified as radio-quiet (R$<$10), exhibit radio powers ranging from 5$\times$10$^{24}$ to 5$\times$10$^{27}$ W Hz$^{-1}$ at 5 GHz, we argue that they could all host a relativistic jet and that orientation could be one of the main causes of the different values of radio-loudness observed in the sample. To support this hypothesis, we show that the number of high-z QSOs detected by EMU is fully consistent with the predictions based on the beaming model, using as reference the high-z blazars discovered at higher radio flux densities. This interpretation requires that, at high redshift, the extended radio emission is intrinsically less powerful, compared to low-z QSOs, either due to the interaction of the jet with the CMB photons and/or because it had not enough time to completely develop, as observed in young radio galaxies. 
   }
   {The results presented here suggest that relativistic jets could have been very common in the early Universe representing up to $\sim$35\% of the luminous (Log$L_{1450\AA}>$46 erg s$^{-1}$) QSO population. 
   Future observations with SKAO will be instrumental to unambiguously confirm the jetted nature of the QSOs currently detected by EMU and to establish if this result still holds when considering optically fainter high-z QSOs.}

   \maketitle
%

\section{Introduction}
The latest generation of telescopes has opened an unprecedented window into the early Universe, allowing us to directly observe the first galaxies during their formation and initial evolution (see \citealt{Harikane2025} for a recent review of the James Webb Space Telescope results). Among these, Active Galactic Nuclei (AGNs) and quasi-stellar objects (QSOs), i.e. galaxies powered by accreting supermassive black holes (SMBHs), offer fundamental insights into this primordial era, providing direct observational constraints on theoretical models of SMBH formation and evolution (e.g.  \citealt{Volonteri2021, Pacucci2021a}). High-redshift AGN are also fundamental to studying the interplay between SMBH activity and galaxy evolution (e.g. \citealt{Vito2022a, Roy2024, Spilker2025}). 

The role of relativistic jets in this picture has yet to be established.
Jets are expected to provide an effective mechanism to explain galaxy feedback (either positive or negative) through the so-called "radio-mode" (e.g. \citealt{Croton2006, Heckman2014, Heckman2023, Hardcastle2020}) and, indeed, direct evidence that this process is active also at high redshift is now emerging (\citealt{Roy2024}). The presence of jets may have an impact also on the evolution of the central SMBH, by slowing down or speeding up its growth: if jets are associated with highly spinning black holes (\citealt{Blandford1977}), the radiative efficiency, $\epsilon$, is expected to be higher than for non‑spinning black holes (\citealt{Thorne1974}). Since the SMBH growth rate scales as (1-$\epsilon$)/$\epsilon$, for a fixed accretion rate (e.g. \citealt{Volonteri2010}), a higher radiative efficiency implies a slower SMBH growth. 
At the same time, the presence of a jet may also speed up the SMBH growth since part of the gravitational energy could be converted into mechanical energy, reducing the total amount of radiation produced by the accretion disk (\citealt{Ghisellini2013a}). As a consequence, for a given optical/UV luminosity (e.g. a QSO radiating at the Eddington luminosity), an accreting SMBH with relativistic jets may grow faster compared to a SMBH without a jet (\citealt{Ghisellini2010, Connor2024}). Both scenarios imply a different accretion history of the two populations and a variation of the jetted/non-jetted QSO space density ratio across the cosmic time. 

From an observational point of view, assessing the actual jetted vs non-jetted space density ratio is not simple,  in particular at high redshift, since it requires a reliable and non-ambiguous way to assess the presence of a relativistic jet in an active nucleus. Historically, radio emission has been used to trace the presence of a synchrotron emitting relativistic jet, either by directly resolving extended structures or by measuring high brightness temperatures in compact sources, through Very Long Baselines Interferometry (VLBI) observations (e.g. \citealt{Blandford2019} and references therein). However, radio emission is also present in non-jetted QSOs, although at lower level, and the distinction between these two classes is not always obvious, particularly at low luminosities  (e.g. \citealt{Panessa2019}). A common way to distinguish between these two classes is the relative radio-to-optical luminosity ratio, called radio-loudness parameter (R\footnote{There are different definitions of the radio-loudness parameter in the literature. Here we adopt the most common one i.e.  R=$\frac{S_{R}}{S_{O}}$, where $S_{R}$ and $S_{O}$ are the k-corrected flux densities at the rest frame frequency 5~GHz and 6.81$\times$10$^{14}$Hz, respectively (corresponding to a wavelength of 4400\AA). This is equivalent to using the monochromatic luminosities at the two frequencies.}). Typically, in QSOs with R$>$10, dubbed as radio-loud, the radio emission is believed to be dominated by relativistic jets. Conversely, QSOs with R$<$10, named radio-quiet, are often, but not always, considered as powered by other mechanisms related, for instance, to synchrotron and thermal Bremsstrahlung emission from star-formation regions.

$3 \times \mathrm{rms}$The fraction of radio-loud QSOs is usually observed to be around 8-10\%, with little or no significant difference from z$\sim$4 up to z$\sim$6 (e.g. \citealt{Banados2015a, Gloudemans2021, Liu2020a, Diana2022, Keller2024}). Considering a wider range of redshifts, however, \citet{Jiang2007} found a significant dependence of the radio-loud fraction both as a function of redshift and of UV luminosity: for a fixed absolute magnitude at 2500\AA\ (-26), the radio-loud fraction decreases from 24.3\% at z=0.5 down to 4.1\% at z=3. Assuming that the value of radio-loudness traces the presence of a relativistic jet, this result suggests that jets were less common in the past compared to what is observed today.

Nevertheless, the reliability of the radio-loudness parameter as a proxy for the presence of a relativistic jet, in particular in objects whose optical/UV emission is obscured,  has been questioned by many authors (e.g. see the discussion in  \citealt{Padovani2016a}). Even considering only unobscured sources, radio-loudness does not seem to provide a clean tool to separate QSOs with/without strong relativistic jets. Using low-frequency (144~MHz) data from LoTSS\footnote{LoTSS stands for LOFAR Two Metre Sky Survey  (\citealt{Shimwell2022}).}, \citet{Macfarlane2021} have shown that a fraction of QSOs whose radio emission is likely dominated by a relativistic jet has R$<$10, reaching values of $\sim$0.1. 
Therefore, limiting the studies to the radio-loud population alone may significantly underestimate the actual number of jetted QSOs.

If we consider the fraction of radio-detected QSOs, not just the radio-loud ones, the results are more critically dependent on the depth of the radio (and optical) data used for the analysis. Interestingly, the LoTSS survey has recently revealed that $\sim$16.5\% of z$>$5 QSOs are radio emitting at low frequency considering $5 \times \mathrm{rms}$ detections (corresponding to an average flux density limit at 144~MHz of $\sim$0.37~mJy/beam), or $\sim$36\% if 
 $2 \times \mathrm{rms}$ 
 detections are considered (0.15~mJy/beam, \citealt{Gloudemans2021}). A possible interpretation is that, at such a low radio emission levels, we are starting to sample a different population of non-jetted QSOs, whose radio emission is due to different processes, like star-formation. Alternatively, these results may suggest that, at high redshift, the fraction of jetted QSOs is significantly higher than previously thought.
 \begin{figure*}
\sidecaption
  \includegraphics[angle=-90,origin=c,width=12cm]{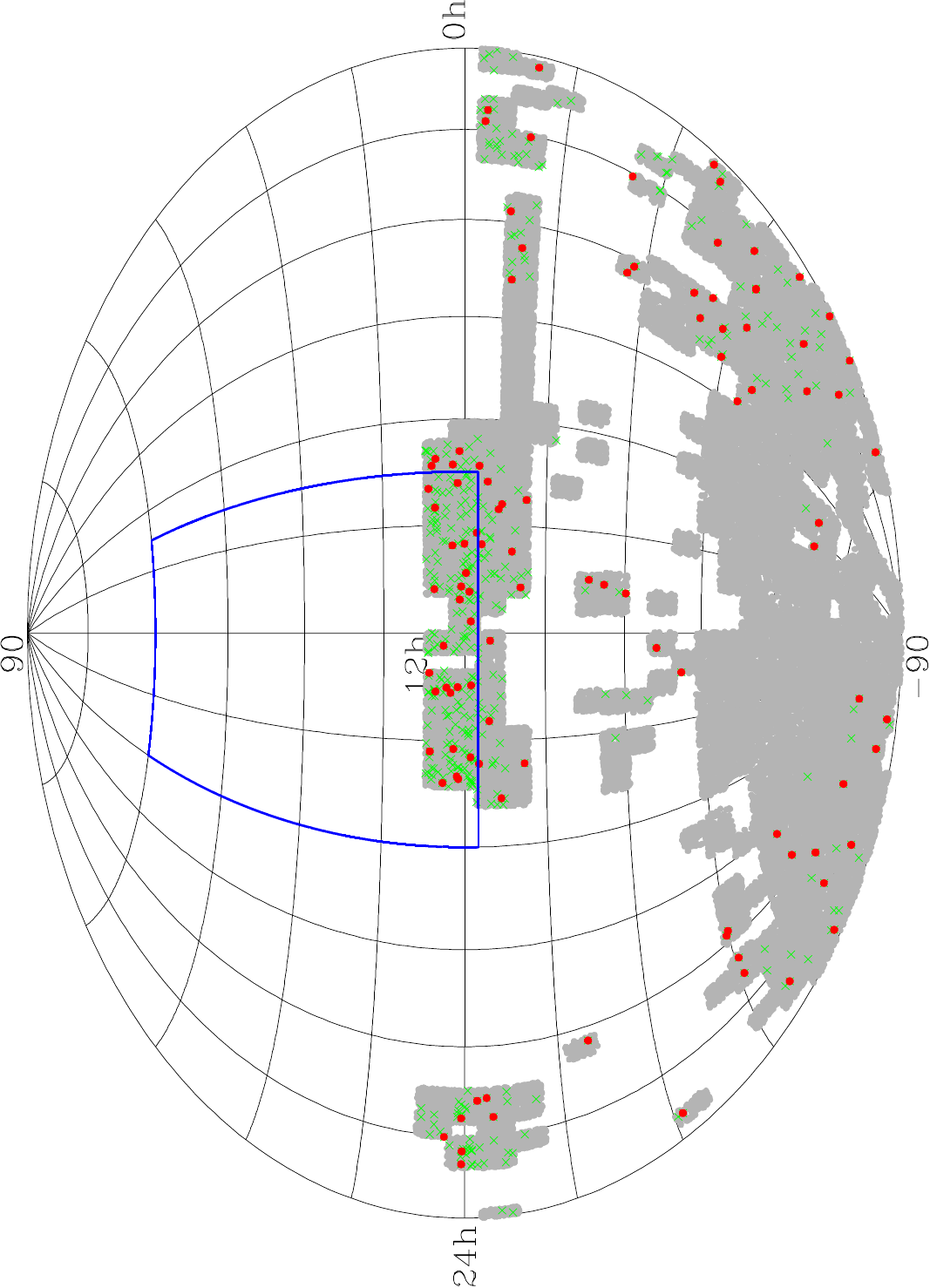}
     \caption{Sky distribution of the z$\geq$4 QSOs (green crosses) falling in the EMU fields (grey areas) up to end of February 2026. Red points represent the objects detected by EMU. The blue line indicates a sky area that is fully covered by SDSS and that is used in Sec.~\ref{sec:pred}.}
     \label{sky_dist}
\end{figure*}

In this paper we first want to extend the result presented in \citet{Gloudemans2021} using a significantly larger sample of QSOs and radio data at much higher frequencies ($\sim$GHz, corresponding to $>$5~GHz for z$>$4 sources) where the relevance of the radio emission from star-formation is expected to be lower (e.g. \citealt{Panessa2019}). For this purpose, we take advantage  of the ongoing Evolutionary Map of the Universe (EMU) survey (\citealt{Norris2011b, Hopkins2025}), which provides one of the deepest looks at $\sim$GHz frequency (0.943~GHz) on a large fraction of sky, to establish the radio-detected fraction of a sizeable list of 808 known z$\geq$4 type~1 QSOs (Section~2, 3 and 4). We further study the high-z QSOs detected in the survey to understand whether the radio emission could be mainly produced by relativistic jets (Section~5). In particular, we use the beaming model (\citealt{Urry1995}) to make predictions on the expected number of misaligned QSOs, using as reference the number of sources whose jet is pointing in our direction (blazars) selected at higher fluxes, and compare them to the observed one. In Section~6 we summarize our conclusions. 

Throughout the paper, we use the mag$_{drop}$ defined as the magnitude in the AB system and corrected for Galactic extinction in the reddest filter of the dropout used to select the sources. This means the {\it r}-filter for 4$\leq$z$<$4.5, the {\it i}-filter for 4.5$\leq$z$<$5.4, the {\it z}-filter for 5.4$\leq$z$<$6.3, and the {\it y}-filter for z$\geq$6.3 objects. By construction, mag$_{drop}$ corresponds to a quite narrow range of rest frame wavelengths between $\sim$1250\AA\ and 1450\AA.  Throughout the paper we assume a flat $\Lambda$CDM cosmology with H$_0$=70 km 
s$^{-1}$ Mpc$^{-1}$, $\Omega_{\Lambda}$=0.7 and $\Omega_{M}$=0.3.  Spectral 
indices are given assuming S$_{\nu}\propto\nu^{-\alpha}$.
    
\section{High-redshift QSOs in The Evolutionary Map of the Universe (EMU)}

The Evolutionary Map of the Universe (EMU, \citealt{Norris2011b, Hopkins2025}) is an ongoing radio survey, conducted with the Australian SKA Pathfinder (ASKAP, \citealt{Hotan2021}) at 0.943~GHz, that will eventually cover the entire southern sky, plus a fraction of the northern sky, with a rms of $\sim$30$\mu$Jy/beam and a resolution of 15$\arcsec$. Up to the end of February 2026 the continuum component catalogues publicly available at CSIRO Data Access Portal\footnote{http://hdl.handle.net/102.100.100/479789?index=1} contain $\sim$10.86 million entries detected ($5 \times \mathrm{rms}$) down to a peak flux density of $\sim$0.15 mJy/beam for a total of $\sim$ 9000 deg$^2$ of sky covered. 
The sky coverage as of the end of February 2026 is shown in Fig.~\ref{sky_dist}.

In order to use the EMU data to establish the fraction of high-z radio-detected QSOs, we have compiled a large sample of 808 z$\geq$4 QSOs from the literature, mostly using NED and Simbad, and from the SDSS spectroscopic database. We have also considered the list of z$>$4 AGN presented in \citet{Perger2017}.
We have then visually checked the optical/NIR spectra from the literature or from SDSS database of each high-z QSO falling in the sky area covered by the EMU catalogue to confirm their redshift and classification as QSO i.e. an AGN with Broad Emission Lines. Ambiguous/uncertain cases are rejected from the sample. If a spectrum is not available, we accept the source only if the discovery paper clearly specifies that it is a QSO with broad emission lines and that the redshift is spectroscopic. In total, we found 808 ``bona-fide" z$>$4 QSOs falling in a sky area covered by EMU at the end of February 2026
(Fig.~\ref{sky_dist}). The majority of these objects have been discovered using SDSS (\citealt{Lyke2020}), DESI (\citealt{Yang2023, Adame2024}), SHELLQ (\citealt{Matsuoka2016, Matsuoka2018}) or SkyMapper (\citealt{Onken2022}) data.
We call this sample HZQ (High-z QSOs).

Notably, 138 objects have 6$\leq$z$\leq$9.3. The presence of deep radio observations at $\sim$1~GHz for such a large set of z$>$6 QSOs offers a unique opportunity to explore the radio properties of high-z QSOs up to the highest values of redshift currently observable. This is of fundamental importance to trace the presence of relativistic jets up to the first epochs of formation and evolution of SMBHs.
  \begin{figure}
   \centering
    \includegraphics[width=8cm, angle=0]{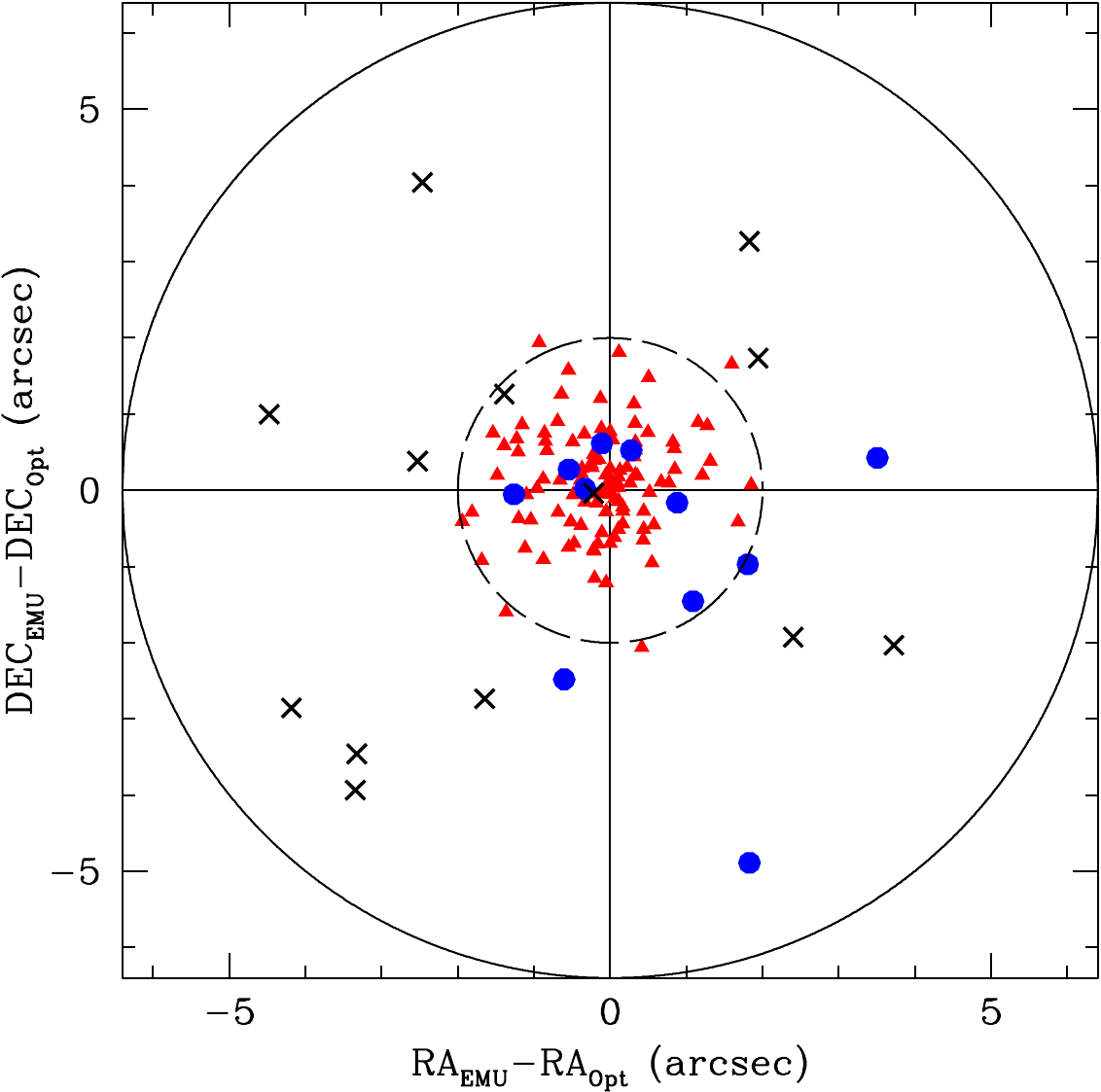}
   \caption{Distribution of the positional offsets between EMU sources and the high-z QSOs counterparts. The offsets in RA include the cos($\delta$) correction. The objects are distinguished on the basis of the integrated-to-peak flux density ratio which is $<$1.2 for the unresolved sources (red triangles) and $>$1.2 for extended sources (blue circles). Black crosses are the matches that have been rejected as likely spurious. The large and small circles have a radius of 6.4$\arcsec$ and 2$\arcsec$ respectively.}
  \label{RA_DEC}
   \end{figure}

\noindent
\begin{center}
    {\it Positional cross-match}
\end{center}

\noindent
The uncertainties of the EMU positions have been evaluated by \citet{Hopkins2025} by cross-matching the EMU sources with either Gaia QSO candidates or with radio sources from the Very Large Array Sky Survey (VLASS, \citealt{Lacy2020, Gordon2021a}). Both comparisons have revealed small systematic offsets of $\sim$0.5$\arcsec$ to the west and $\sim$0.1$\arcsec$ to the north, respectively, of unknown origin. 
Given the small values of these offsets and the fact that their origin is not clear, we decided to ignore them in the following analysis.
The standard deviations on $\Delta\alpha$ and $\Delta\delta$ are larger when Gaia QSOs/EMU are considered ($\sim$2.1$\arcsec$) and lower in the VLASS/EMU comparison ($\sim$0.68-0.76$\arcsec$). It is possible that the larger offset observed with the Gaia QSO candidates is in part related to the presence of complex radio morphologies (e.g. lobe-dominated QSOs) that may cause an intrinsic radio/optical offset, not related to the astrometric errors. Since we want to be as complete as possible, including also QSOs with complex radio morphologies, we start from the standard deviations measured with Gaia QSO. Assuming Gaussian distributions and a standard deviation on $\Delta\alpha$ and $\Delta\delta$ of 2.1$\arcsec$, the positional tolerance required to achieve a $\sim$99\% completeness level is 6.4$\arcsec$. 
Using this positional tolerance, we found 137 unique matches between our list of high-z QSOs and the EMU catalogue of components.
The distribution of the radio/optical positional offset is reported in Fig.~\ref{RA_DEC}.
   \begin{figure}
   \centering
    \includegraphics[width=8cm, angle=0]{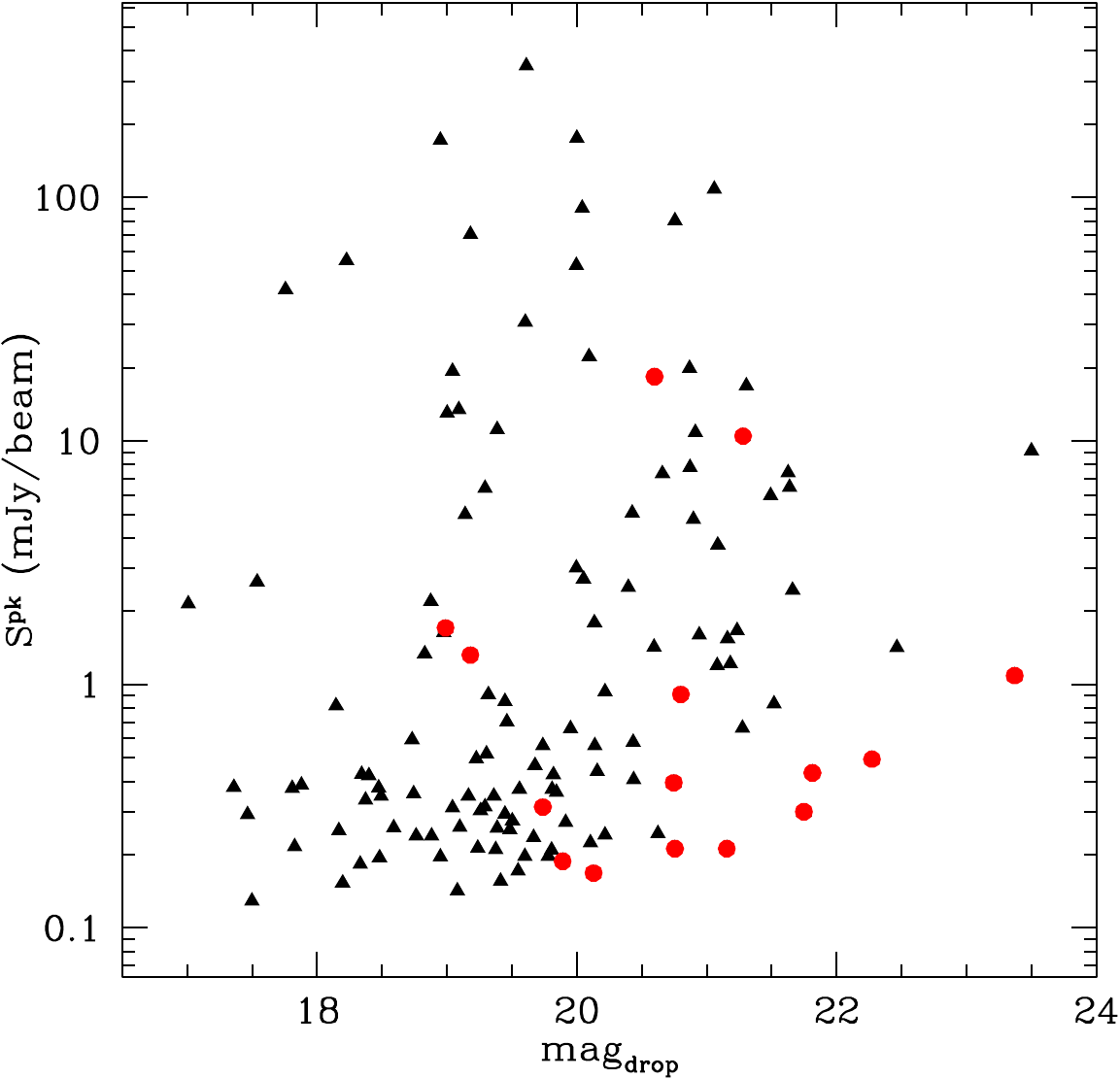}
    \caption{EMU peak flux density versus mag$_{drop}$. Black triangles are QSOs with 4$\leq$z$<$5.5 while red points are QSOs with 5.5$\leq$z$\leq$6.6.
    }
    \label{radio_vs_mag}
    \end{figure}
   \begin{figure}
   \centering
    \includegraphics[width=6cm, angle=0]{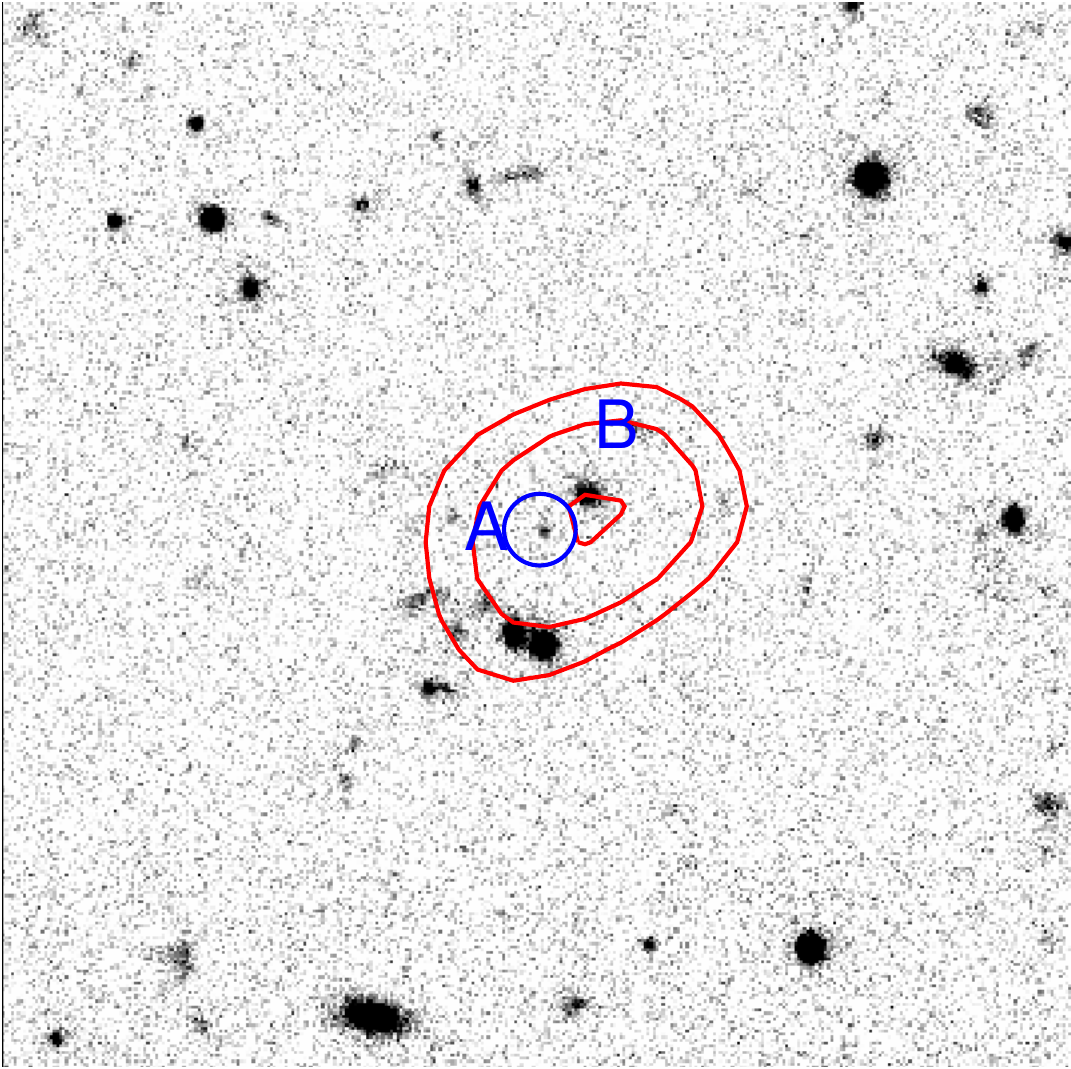}
    \caption{Example of a likely spurious radio-optical match that has been removed from the sample. The most probable counterpart of the radio source (red contours) is the bright object close to the radio peak emission (B) and not the high-z QSO (HZQJ120754.1-000553, A), a z=6.1 QSO discovered by \citet{Matsuoka2016} indicated by the blue circle. The 1$\arcmin\times$1$\arcmin$ image is in the y-filter and is taken from the Subaru High-z Exploration of Low-Luminosity Quasars (SHELLQs, \citealt{Matsuoka2016}) project. Radio (EMU) contours are plotted by steps of $\sqrt{2}$ starting from $3 \times \mathrm{rms}$ (0.034~mJy/beam).
    }
    \label{spurious}
    \end{figure}


\begin{figure*}
\centering
\includegraphics[width=18cm, angle=0]{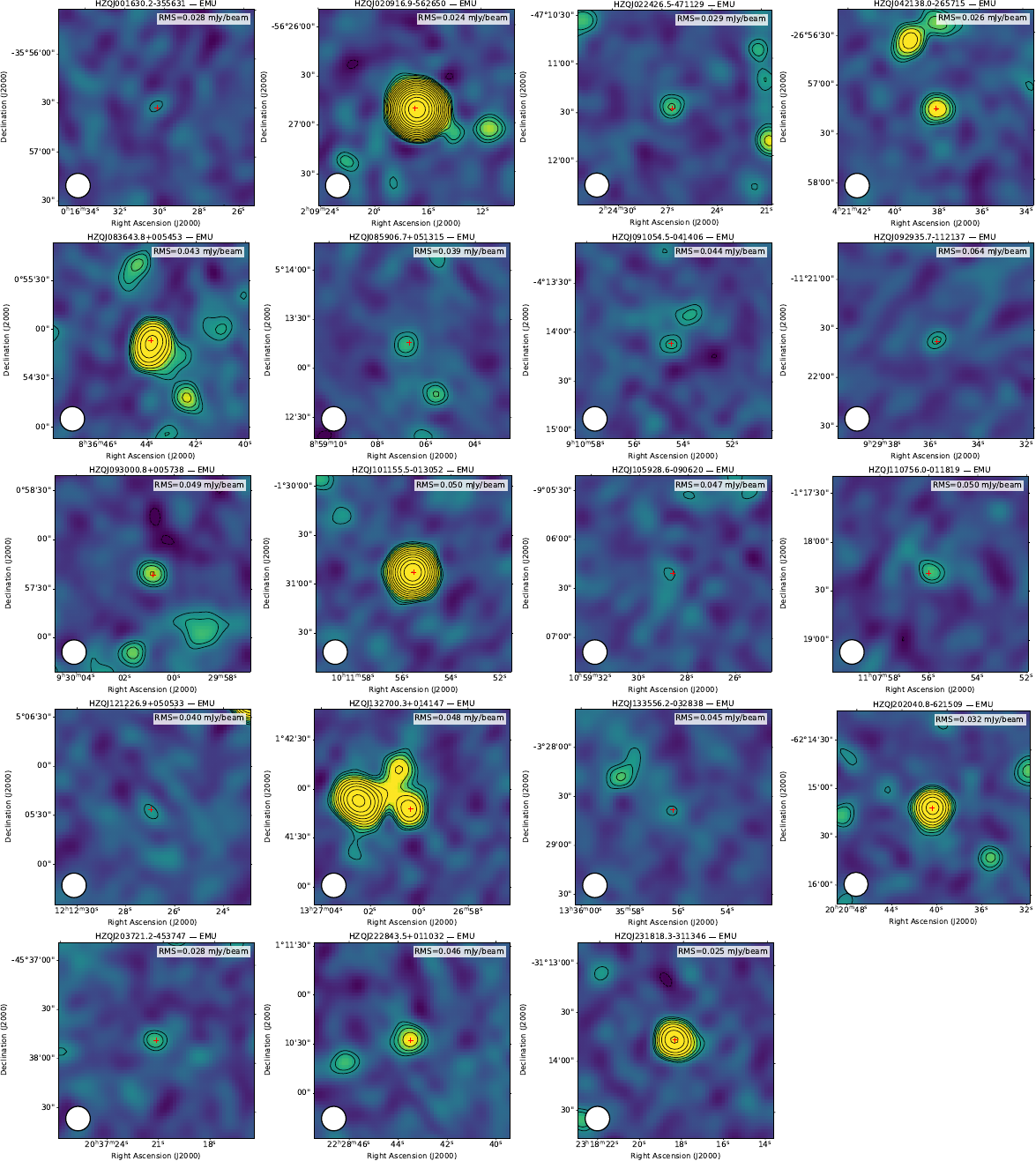}
\caption{2$\arcmin\times$2$\arcmin$ Radio maps at 0.943~GHz of the z$>$5.5 QSOs detected in EMU, including those tentatively detected with a significance between 3 and 
$5 \times \mathrm{rms}$, not present in the EMU catalogue. Contour levels are plotted by steps of $\sqrt{2}$ starting from 
$3 \times \mathrm{rms}$
(reported in the inset). The red cross indicates the optical position of the high-z QSO. The beam size is reported in the bottom-left corner. The maps of all the z$\geq$4 QSOs detected by EMU is reported in \href{https://doi.org/10.5281/zenodo.24462937}{Appendix~A}}
\label{maps_zgt5_5}
\end{figure*}
   \begin{figure}
   \centering
   \includegraphics[width=6cm, angle=0] 
   {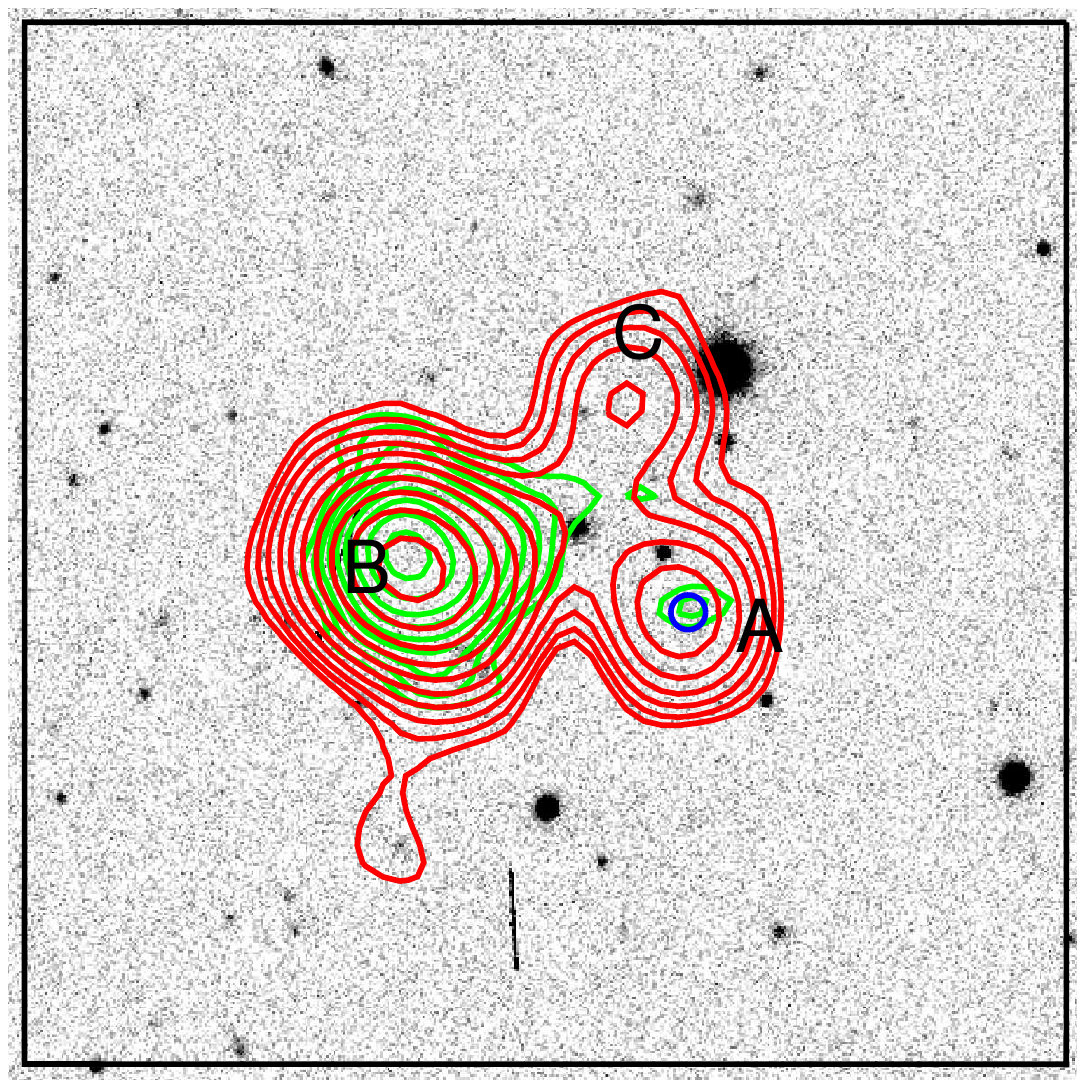} 
    \caption{Overlay of the radio (EMU, red lines and LoTSS, green lines) contours on DECaLS z-filter images (2$\arcmin\times$2$\arcmin$) for HZQJ132700.3+014147 at z=6.23 with a complex radio morphology. Contour levels are plotted by steps of $\sqrt{2}$ starting from 0.048~mJy/beam, (EMU) and 0.7~mJy/beam (LoTSS).
    The position of the high-z QSO is indicated by the blue circle.
    }
    \label{overlays}
    \end{figure}

We have then assessed the expected number of spurious matches by off-setting the positions by several arcminutes along different directions. We found that about eight matches could be just spurious coincidences.
In order to find at least some of these possible spurious matches, we have analyzed all the radio maps of the sources, with particular attention to the cases with a large radio/optical positional offset. The overlay of the radio contours on the optical image has revealed 13 cases where the radio emission is probably not related to the high-z QSO (an example is reported in Fig.~\ref{spurious}, left panel). 
As expected, a significant fraction of the matches with a large ($>$2$\arcsec$) radio/optical positional offset (19 objects) are either spurious (11) or extended (4). 
Since the number of discovered spurious matches is consistent with the expected one, we believe that the resulting sample, after the removal of the possible spurious matches, is very clean. 

Finally, we searched for optical and near infrared photometric data using Pan-STARSS1 (PS1, \citealt{Chambers2016,Flewelling2020}) and  DELVE (\citealt{Drlica-Wagner2022}). In the few cases where a magnitude was not available in these catalogues, we used the values published in the discovery papers. Most of the collected  magnitudes come from DELVE. 
After correcting all the magnitudes for the Galactic extinction, we have computed the $mag_{drop}$ i.e. a  magnitude in a redshift-dependent filter corresponding to similar rest frame wavelengths ($\sim$1250-1450\AA) for all the objects.  

A table containing the full list of 124 high-z QSOs detected by EMU is published electronically at the CDS. Their radio flux densities and magnitudes are plotted in Fig.~\ref{radio_vs_mag} while all the radio maps can be found in \href{https://doi.org/10.5281/zenodo.24462937}{Appendix~A}.

\section{Radio-emitting QSOs at z$>$5.5}
\begin{table*}
\scriptsize
 \centering
 \caption{z$>$5.5 QSOs detected by EMU}
 \label{tab:sample_zgt5_5}
 \begin{tabular}{lrrccrrrll}
  \hline
name                 &  RA        &   DEC      &  z     & mag$_{drop}$  & $S_{0.943GHz}^{pk}$ & log R & d    & Ref. Id & Ref. radio \\
                     &  (J2000)   &   (J2000)  &        &      & (mJy/beam)         &       & ('') &         &            \\
  \hline
    \hline
HZQJ001630.2$-$355631 $^*$ &   4.12608 & -35.94206 & 6.18 & 20.87$\pm$ 0.03 (z              R) &     0.10$\pm$    0.03 &  0.6 &  1.50 & \citet{Belladitta2025}&this work\\
HZQJ020916.9$-$562650 &  32.32053 & -56.44734 & 5.61 & 20.60$\pm$ 0.03 (z              D) &    18.40$\pm$    0.03 &  2.9 &  1.34 & \citet{Ighina2024k}$^1$&\citet{Ighina2024k}$^1$\\
HZQJ022426.5$-$471129 &  36.11057 & -47.19149 & 6.51 & 19.89$\pm$ 0.05 (y              R) &     0.19$\pm$    0.03 &  0.3 &  1.56 & \citet{Reed2017}&this work\\
HZQJ042138.0$-$265715 &  65.40854 & -26.95434 & 6.14 & 19.74$\pm$ 0.01 (z              D) &     0.31$\pm$    0.03 &  0.6 &  0.73 & \citet{Banados2016}&this work\\
HZQJ083643.8$+$005453 & 129.18275 &   0.91479 & 5.77 & 18.99$\pm$ 0.01 (z              D) &     1.71$\pm$    0.03 &  1.1 &  2.55 & \citet{Fan2001b}&\citet{Fan2001b}\\
HZQJ085906.7$+$051315 & 134.77792 &   5.22104 & 5.75 & 21.15$\pm$ 0.03 (z              P) &     0.21$\pm$    0.02 &  1.1 &  2.03 & \citet{Yang2023}&this work\\
HZQJ091054.5$-$041406 & 137.72721 &  -4.23522 & 6.63 & 20.75$\pm$ 0.13 (y              P) &     0.21$\pm$    0.04 &  0.7 &  0.28 & \citet{Wang2019}&this work\\
HZQJ092935.7$-$112137 & 142.39904 & -11.36044 & 5.61 & 20.75$\pm$ 0.04 (z              D) &     0.40$\pm$    0.10 &  1.0 &  0.72 & \citet{Banados2022}&this work\\
HZQJ093000.8$+$005738 & 142.50354 &   0.96067 & 5.92 & 21.81$\pm$ 0.14 (z              D) &     0.43$\pm$    0.06 &  1.6 &  0.93 & \citet{Matsuoka2019a}&this work\\
HZQJ101155.5$-$013052 & 152.98142 &  -1.51464 & 5.54 & 21.28$\pm$ 0.05 (z              D) &    10.47$\pm$    0.07 &  2.7 &  0.06 & \citet{Ighina2024k}&\citet{Ighina2024k}\\
HZQJ105928.6$-$090620 $^*$ & 164.86921 &  -9.10567 & 5.92 & 20.92$\pm$ 0.02 (z              D) &     0.16$\pm$    0.05 &  0.8 &  2.60 & \citet{Willott2010a}&this work\\
HZQJ110756.0$-$011819 & 166.98338 &  -1.30528 & 6.06 & 21.75$\pm$ 0.09 (z              D) &     0.30$\pm$    0.05 &  1.5 &  1.71 & \citet{Matsuoka2022}&this work\\
HZQJ121226.9$+$050533 $^*$ & 183.11242 &   5.09264 & 6.44 & 19.98$\pm$ 0.05 (y              P) &     0.15$\pm$    0.04 &  0.6 &  2.50 & \citet{mazzucchelli2017}&this work\\
HZQJ132700.3$+$014147 & 201.75146 &   1.69658 & 6.23 & 23.37$\pm$ 0.08 (y              R) &     1.09$\pm$    0.05 &  2.6 &  0.74 & \citet{Matsuoka2022}&this work\\
HZQJ133556.2$-$032838 $^*$ & 203.98433 &  -3.47728 & 5.67 & 18.89$\pm$ 0.01 (z              D) &     0.15$\pm$    0.05 & -0.1 &  1.00 & \citet{Yang2017}&this work\\
HZQJ202040.8$-$621509 & 305.17019 & -62.25255 & 5.70 & 19.18$\pm$ 0.01 (z              D) &     1.32$\pm$    0.03 &  0.8 &  0.09 & \citet{Wolf2024}&\citet{Wolf2024}\\
HZQJ203721.2$-$453747 & 309.33858 & -45.62986 & 5.64 & 20.13$\pm$ 0.02 (z              D) &     0.17$\pm$    0.04 &  0.1 &  0.82 & \citet{Yue2021}&this work\\
HZQJ222843.5$+$011032 & 337.18125 &   1.17561 & 5.95 & 22.27$\pm$ 0.12 (z              D) &     0.49$\pm$    0.04 &  1.9 &  0.69 & \citet{Zeimann2011}&\citet{Zeimann2011}\\
HZQJ231818.3$-$311346 & 349.57646 & -31.22954 & 6.44 & 20.80$\pm$ 0.08 (y              R) &     0.91$\pm$    0.02 &  1.8 &  0.19 & \citet{Decarli2018a}&\citet{Ighina2021a}\\
\hline
 \end{tabular}
\tablefoot{Column 1: Source's name. The asterisk indicates that the source is not present in the catalogue but has been tentatively detected at 
$3 \times \mathrm{rms}$; Column 2 and 3: radio (EMU) position; Column 4: redshift; Column 5: mag$_{drop}$ (the filter is indicated between parentheses together with the survey: P=PS1, D=DELVE, R=reference in the Ref. Id column); Column 6: peak flux density at 0.943GHz from EMU catalogue; Column 7: 
 log of radio-loudness calculated using the radio and optical k-corrected flux densities at 5~GHz and at 6.81$\times$10$^{14}$ Hz (corresponding to $\lambda$=4400\AA). We assume a power law for both the radio and the optical spectral shape, using a computed radio slope (if available) or assuming $\alpha_r$=0.7. To compute the k-corrected flux density at 4400\AA, we start from the near-infrared magnitude at 3.4~micron (W1) from the Wide-field Infrared Survey Explorer (NEOWISE; \citealt{Mainzer2014}), which provides the photometric point closest to the 4400\AA\ rest frame wavelength for z$\geq$4 sources, and we use the optical spectral index computed between W1 and the z-filter (or y-filter for z$>$6.3 sources). If W1 was not available, we computed R using mag$_{drop}$ and assuming $\alpha_0$=0.44 (\citealt{VandenBerk2001});
 Column 8: distance between the EMU source and the optical position of the QSO; Column 9: reference to the spectroscopic identification ($^1$=also in \citealt{Wolf2024}); Column 10: reference where the radio detection is discussed for the first time ($^1$=also in \citealt{Wolf2024}) The entire sample of z$\geq$4 QSOs detected by EMU is reported in electronic form at the CDS.}
\end{table*}

The discovery of radio-emitting QSOs at z>5.5 is of great interest since the luminosity function of jetted QSOs is largely unconstrained at these redshifts, with the most recent estimates limited at z$\leq$5 (e.g. \citealt{Slaus2024, Wang2026}). Moreover, z$>$5.5 QSOs give the unique opportunity of tracing the neutral hydrogen at the end of the re-ionization epoch through the detection of the red-shifted 21 cm HI line in absorption along the line-of-sight of these high-z radio sources (the 21-cm forest, e.g. \citealt{Soltinsky2025}).

Among the 124 z$\geq$4 QSOs detected by EMU there are 15 objects at 5.5$\leq$z$\leq$6.63.  Six of these are already known as radio emitters, while the remaining nine are detected for the first time at these wavelengths. 
Since the sources in the EMU catalogue are detected using a relatively high detection threshold (5$\times  rms$), we have searched for additional detections among the z$>$5.5 QSOs, at lower significance, i.e. between 3 and 5$\times rms$. To this end, we have looked for a peak emission brighter than 3 times the local rms at the optical position of the QSO.
We have found 4 possible new detections of QSOs at z$>$5.5. 
We have repeated the same analysis using randomly shifted positions, and we did not find any significant excess, supporting the reliability of these new detections. 

Considering also these 4 detections, we have a total of 19 z$>$5.5 QSOs seen by EMU down to $\sim$0.1 mJy/beam, 13 of which detected in the radio for the first time. The properties of these 19 QSOs are reported in table~\ref{tab:sample_zgt5_5} while their radio maps are shown in Fig.~\ref{maps_zgt5_5}.

Among the newly discovered radio-emitters there is a potentially interesting object,  HZQJ132700.3+014147 at z=6.23 (\citealt{Matsuoka2022}), whose radio map (Fig.~\ref{overlays}) reveals the presence of many nearby radio sources/components.
The overlay of the radio contours on the optical image shows no clear counterparts coincident with the other two radio components to the east (B) and north (C), respectively. 
Radio components A and B are both detected at 144~MHz by LoTSS with an integrated flux density of 0.95~mJy and 20.3~mJy respectively. The (non-simultaneous) spectral index of the radio component A, positionally coincident with the high-z QSO, is very flat ($\alpha_R$=-0.07) while component B has a steep radio spectrum ($\alpha_R$=0.85). Source B is also clearly extended in both EMU and LoTSS, in agreement with the observed steep radio spectrum.
With these data, it is difficult to establish if components B and C are related to the high-z QSO. 
It is possible that these are the radio lobes produced by the bright central object (an elliptical galaxy at z=0.53). 
Considering only the radio component coincident with the optical position,  HZQJ132700.3+014147 has a very high radio-loudness (R=380$^{+260}_{-170}$), due to its faint NIR magnitude (y$_{AB}$=23.37, \citealt{Matsuoka2022}). To date, only two z$>$6 QSOs have R$>$300 (\citealt{Belladitta2020b, Gloudemans2022}) and they are either confirmed or good candidate blazars. It is therefore possible that also J132700.3+014147 is a blazar. This is further supported by the very flat radio spectrum observed between 0.144 and 0.943~GHz (corresponding to 1.0-6.8~GHz in the source's rest-frame) If confirmed, this would be one of the very few blazars discovered so far at z$>$6 (three in total,  \citealt{Belladitta2020b, Banados2025, Ighina2025f}). Even more importantly, it would be the first blazar at these redshifts found at relatively low radio (P$_{1.4GHz}<$10$^{26}$ W Hz$^{-1}$) and UV (L$_{1450\AA}<10^{46}$ erg s$^{-1}$) luminosities. The observed UV luminosity (L$_{1450\AA}=2\times10^{45}$ erg s$^{-1}$, \citealt{Matsuoka2022}) is suggestive of a central SMBH with a small mass, of the order of 10$^7$-10$^8$ M$_{sun}$, which is significantly lower than those found, to date, in high-z QSOs detected in the radio (typically above 10$^9$ M$_{sun}$).
This object shows the potential of EMU for exploring the still unknown faint part of the radio/optical luminosity function of blazars at high redshift.

Finally, it is worth noting that in the sky area currently covered by EMU there are also two broad line AGNs at z$>$7: HSC J124353+010038 at z=7.07 (\citealt{Matsuoka2019}) and the ``Little Red Dot" CAPERS-LRD-z9 (HZQJ100032.7+021828), at z=9.288 recently discovered by \citet{Taylor2025}. None of these AGNs is detected by EMU, implying a $3\times \mathrm{rms}$ upper limit on the 0.943GHz flux density of 0.14 and 0.13 mJy/beam respectively. Given the optical faintness of these two objects (Y=23.57 and F150W=29.12, respectively), the upper limits on the radio flux densities correspond to very loose limits on the radio-loudness parameter (R$<$80 and R$<$15000, respectively) that do not exclude a radio-loud nature.

\section{The fraction of radio-detected QSOs}
The 124 high-z QSOs that match a radio source correspond to 15\% of the total number of objects falling in the EMU fields (808 in total)
\footnote{We note that, while the majority (110/124) of the high-z QSOs in the sample is optically selected, a small fraction was discovered starting from a radio survey. This may slightly affect the computed fraction of radio-detected QSOs discussed in this section. However, the impact of this bias is only marginal: if we exclude the radio selected sources we obtain a fraction of 14\% instead of 15\%. Since also the exclusion of these objects is expected to introduce a bias, we keep them in the analysis.}. 
This fraction, in itself, is difficult to interpret as the starting list of high-z QSOs is very heterogeneous, containing both very faint objects detected in deep optical/NIR surveys and relatively bright sources coming from shallow surveys like the SDSS. 
Since the probability of exceeding the EMU flux-density limit is a joint function of optical flux and radio loudness, the total fraction of objects detected by EMU inevitably reflects the different depths of the optical/IR surveys used to select the sources.
   \begin{figure}
   \centering
    \includegraphics[width=7cm, angle=0]{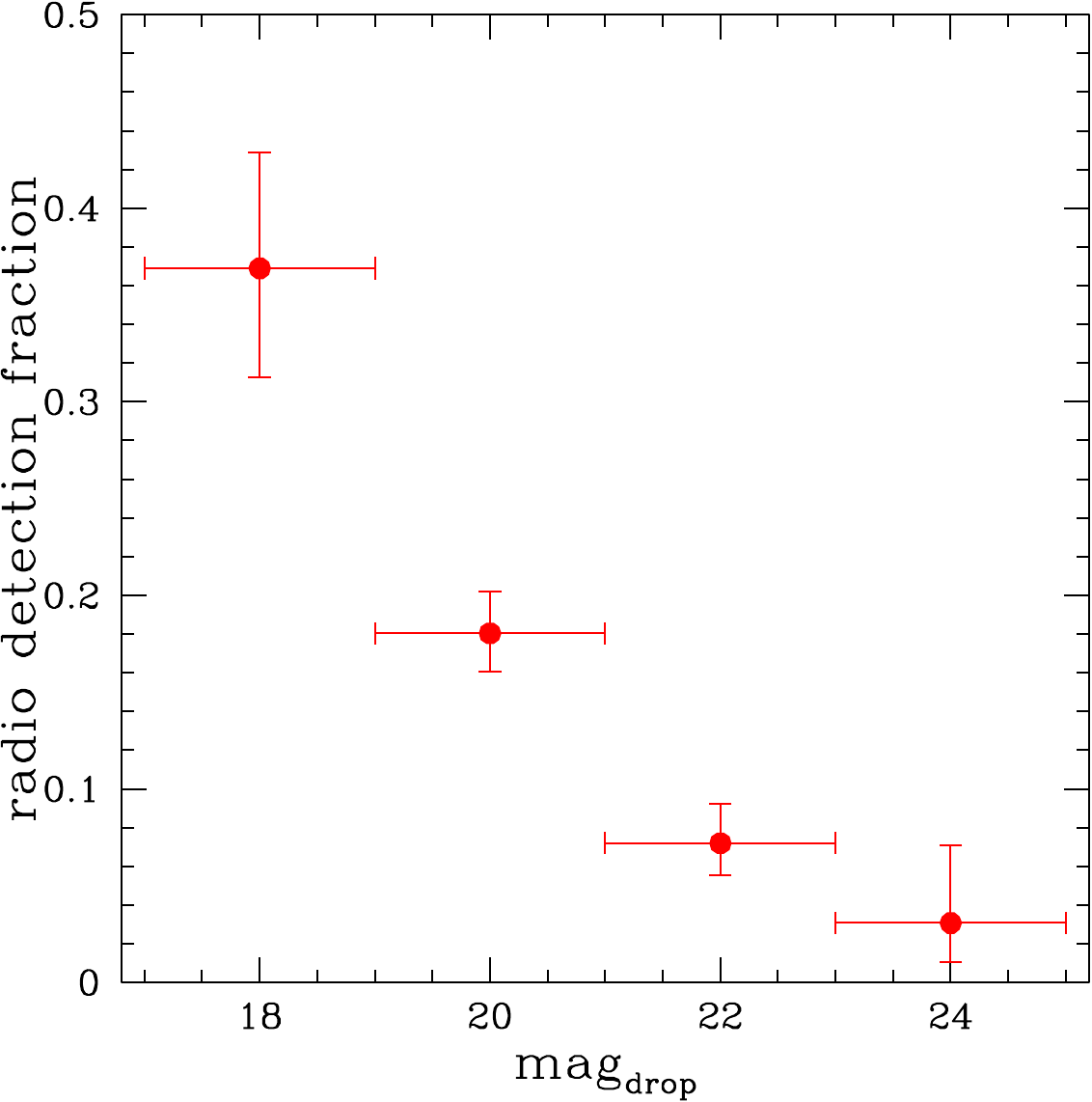}
    \caption{EMU detection fraction of high-z QSOs as a function of mag$_{drop}$.
    Vertical error bars are the Poissonian errors at 68\% confidence level.} 
              \label{fig_frac_vs_mag}
    \end{figure}
To account for this effect, in Fig.~\ref{fig_frac_vs_mag} we show the EMU detection fraction in bins of mag$_{drop}$. 
As is clear from the figure, there is a strong dependence of this fraction with the source's optical brightness. The detection fraction ranges from $\sim$3\% at mag$_{drop}\sim$24 up to $\sim$37\% for mag$_{drop}\sim$18. The value observed in the total sample (15\%) reflects the fact that most of the sources come from relatively shallow surveys, like the SkyMapper (\citealt{Onken2022, Wolf2020}) and SDSS (\citealt{Lyke2020}), that have magnitude limits of $\sim$19-21. 

\begin{figure}
   \centering
    \includegraphics[width=9cm, angle=0]{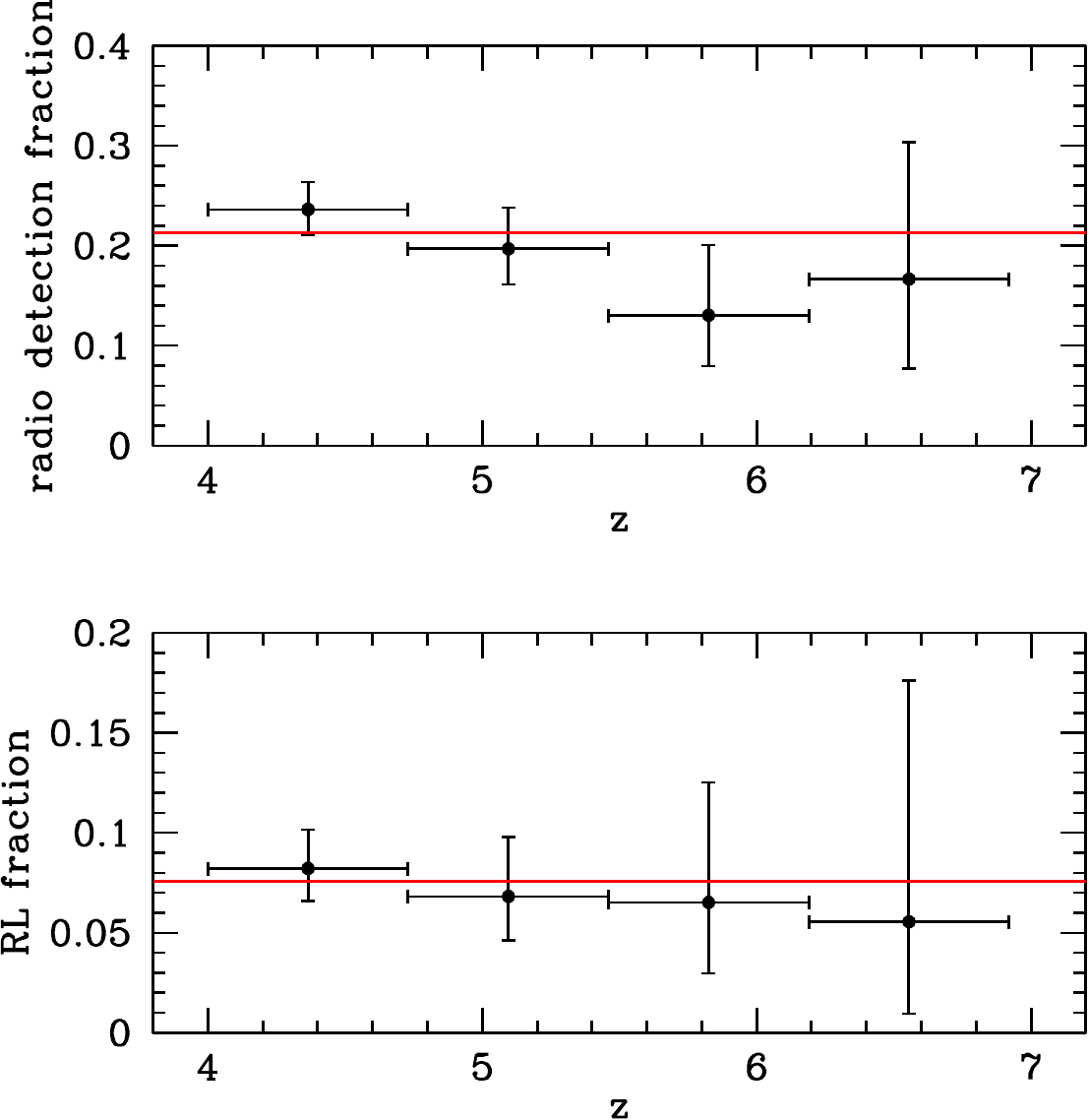}
    \caption{{\it Upper Panel:} Fraction of radio detected high-z QSOs (with mag$_{drop}\leq$21) present in the sky area currently covered by EMU vs redshift; {\it Lower Panel:} Fraction of high-z QSOs (mag$_{drop}\leq$21) with R$>$10 vs redshift. The red lines indicate the average values. Vertical error bars as in Fig.~\ref{fig_frac_vs_mag}}
              \label{R_frac}.
    \end{figure}

In Fig.\ref{R_frac} (upper panel) we show the fraction of high-z QSOs  detected in EMU as a function of z
at a fixed magnitude limit (mag$_{drop}\leq$21). At such a magnitude limit, there are 486 z$\geq$4 QSOs falling in the EMU fields, 104 of which are detected and 36 are radio-loud (R$\geq$10). 
In the lower panel of Fig.\ref{R_frac}, instead, we show the fraction of radio-loud QSOs with the same magnitude limit. 
In the two panels we do not see any significant dependence with redshift. At these magnitudes (mag$_{drop}\leq$21) the fraction of radio detected QSOs (21.3\%) is almost three times the fraction of radio loud QSOs (7.4\%) computed in the same sample.

\section{Jetted or non-jetted QSOs?}
\begin{figure}
   \centering
    \includegraphics[width=9cm, angle=0]{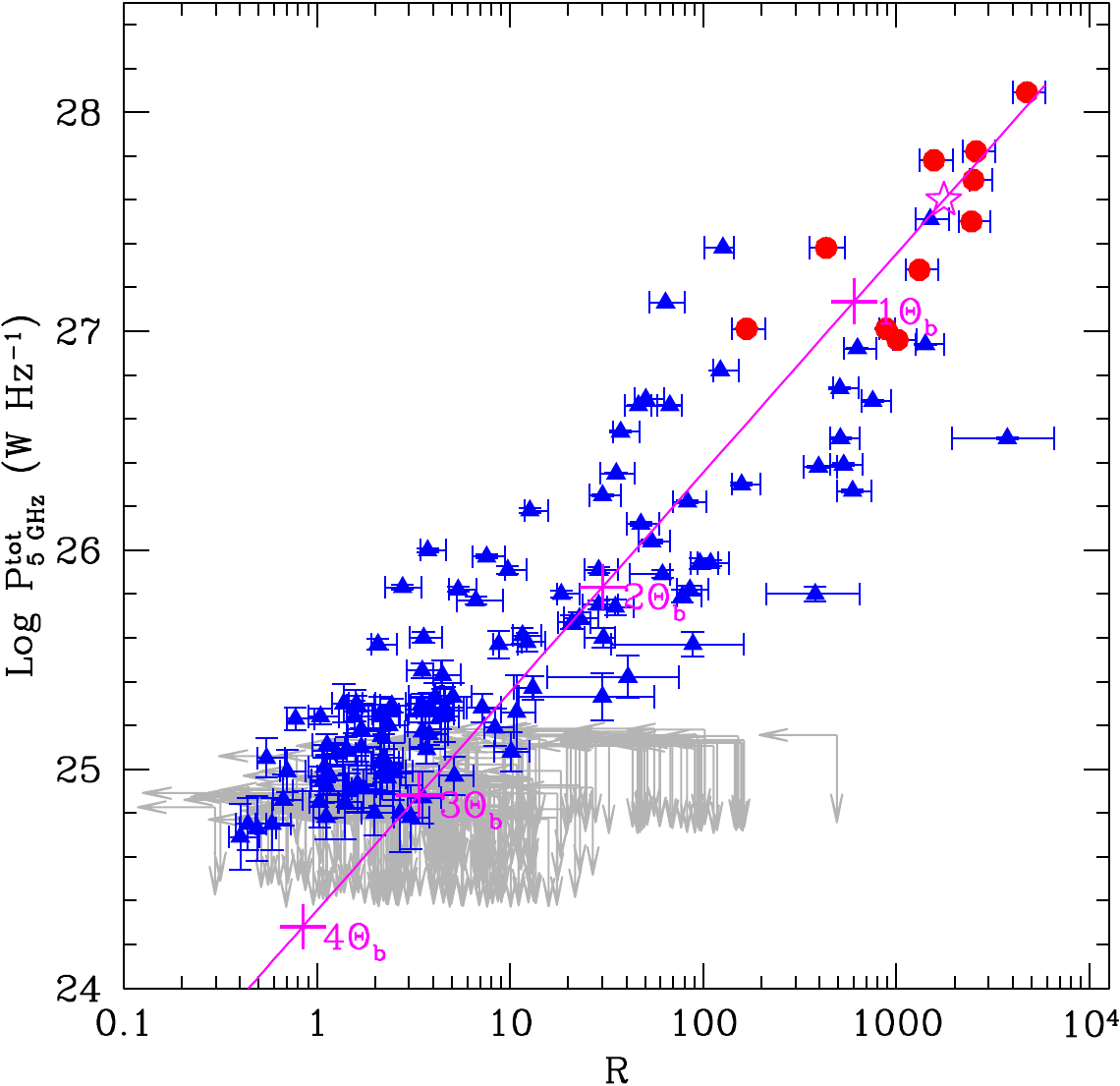}
    \caption{Radio power at 5~GHz (in the source's rest-frame)  versus radio-loudness parameter of the z$>$4 QSOs currently detected in EMU. Red points indicate the sources currently classified as blazars. Grey arrows represent the sources not detected in EMU. Error-bars reflect both the statistical errors on flux densities and the uncertainties due to the unknown radio/optical spectral slopes. The solid magenta line shows the expected values of P$_{5GHz}$ and R according to the beaming model for different viewing angles (in units of $\theta_b$. $\theta_b\sim$5.7 deg for $\Gamma$=10), assuming f$_5$=1 and using as reference the average value of P$_{5GHz}$ and R computed for all the 10 blazars present in the sample (indicated by the empty star). See text for more details}
              \label{lum_R}
    \end{figure}

In the previous sections we have seen that the fraction of radio-detected high-z QSOs can be as high as 35\%, depending on the optical magnitude.  This is likely a lower limit since deeper radio observations would probably detect more objects. The fact that the fraction of radio detected QSOs is much larger than the fraction of QSOs formally classified as radio-loud may not be surprising since it is well known that all QSOs, including the radio-quiet ones, emit at radio wavelengths at some level (e.g. \citealt{Panessa2019}) and, therefore, the detection fraction is going to progressively increase with the increasing of sensitivity of the radio surveys. Indeed, the relatively low values of radio-loudness (R$<$10) of many high-z QSOs detected in EMU (see Fig.~\ref{lum_R}) seem to indicate that, thanks to the depth of the radio data, we are now detecting, besides the usual jetted QSOs, a new population of QSOs, powered by different mechanisms, like star-formation.

However, if we consider the radio power of the QSOs detected in EMU (Fig.~\ref{lum_R}) we see that the values range from 5$\times$10$^{24}$ to 5$\times$10$^{27}$ W Hz$^{-1}$ at 5~GHz (in the source's rest frame\footnote{Since the observing frequency of 0.943~GHz corresponds to 4.7-7.1~GHz in the source's rest frame for 4$\leq$z$\leq$6.5, we decided to use the radio power at 5~GHz (P$_{5GHz}$), which is a common frequency studied in the literature, to limit the uncertainties due to the unknown radio spectral index. As expected, the values of P$_{5GHz}$ change - on average - only by 6\% when very different values of $\alpha_R$ are assumed, from $\alpha_R$=0 to $\alpha_R$=2.}) i.e. well within the typical range of radio power observed in the radio-loud population (\citealt{Padovani2016a}). On the contrary, the population of radio-quiet AGNs cover radio powers much lower, typically between 10$^{18}$ and 10$^{24}$ W Hz$^{-1}$ (Ulvestad \& Wilson 1989; Ulvestad \& Ho 2001a).
Similarly, based on a large sample of QSOs detected by LoTSS, \citet{Macfarlane2021} have shown that, for radio powers at 150~MHz larger than $\sim$5$\times$10$^{25}$ W Hz$^{-1}$, corresponding to P$>$4$\times$10$^{24}$ W Hz$^{-1}$ at 5~GHz ($\alpha_R$=0.7), the majority of the radio detected QSOs are dominated by relativistic jets. 

The origin of radio emission in RQ AGNs is still debated (\citealt{Panessa2019}), but there are some indications that it could be related to star-formation, as suggested, for instance, by the observed similarities with the starburst galaxy population in terms of L$_{RADIO}$--L$_{FIR}$ relation (see \citealt{Padovani2016a} and references therein).
On the basis of the relation between radio and star-formation rate (SFR, e.g. 
\citealt{Condon2002}) we estimate that the radio powers observed in the high-z QSOs detected by EMU, including those objects with low radio-loudness values, would require SFR of the order or above 10$^4$ M$_{sun}$ y$^{-1}$ to be explained in nearly all cases. While the exact values of SFR depend on the adopted radio–SFR calibration and its possible redshift evolution, overall they seem too high to be explained only by star-formation, even considering high-z sources (e.g. \citealt{Popesso2023, Pouliasis2022, Mazzucchelli2025}) and/or very luminous QSOs (\citealt{Macfarlane2021}).
For this reason, it seems quite unlikely that the dominant mechanism for the production of the radio luminosity in the high-z QSOs detected by EMU is star-formation although we cannot exclude that this process can contribute, at some level, to the observed radio emission of individual sources.

Another possibility discussed in the literature  to explain the radio emission of radio-quiet AGNs is related to the hot corona that is also responsible for the X-ray luminosity (\citealt{Panessa2019}). This hypothesis was supported by the significant correlation found by \citet{Laor2008} between the X-ray (0.2-20~keV) and the radio (5~GHz) emission of the Palomar-Green sample of AGNs that follows remarkably well the relation observed in coronally active stars. Using this relation (L$_R$/L$_X\sim$10$^{-5}$) we can estimate the expected X-ray emission of the high-z QSOs with R$<$10 detected in EMU under the assumption that the observed radio luminosity is due to the corona. We obtain X-ray fluxes between 10$^{-13}$ and 10$^{-12}$ erg s$^{-1}$ cm$^{-2}$ that should allow a clear detection in the eROSITA Sky Survey (eRASS1, \citealt{Merloni2024}). Instead, only 7 objects out of the 99 falling in the sky area covered by eRASS1 are detected and, also in these cases, the observed flux is more than one order of magnitude fainter than the predicted one. We conclude that the observed radio emission is too high to be produced by the corona, even for the objects of the sample formally classified as radio-quiet. 

While we cannot exclude that the high-z QSOs detected by EMU with R$<$10 are powered by a different mechanism with respect to the most luminous ones, in the next sections we want to explore the hypothesis that the radio emission of all (or most of) the QSOs in the sample discussed here, including those formally classified as radio-quiet, is produced by a relativistic jet and that orientation is the main reason for the low values of radio-loudness observed in some objects. 

   \begin{figure}
   \centering
    \includegraphics[width=9cm, angle=0]{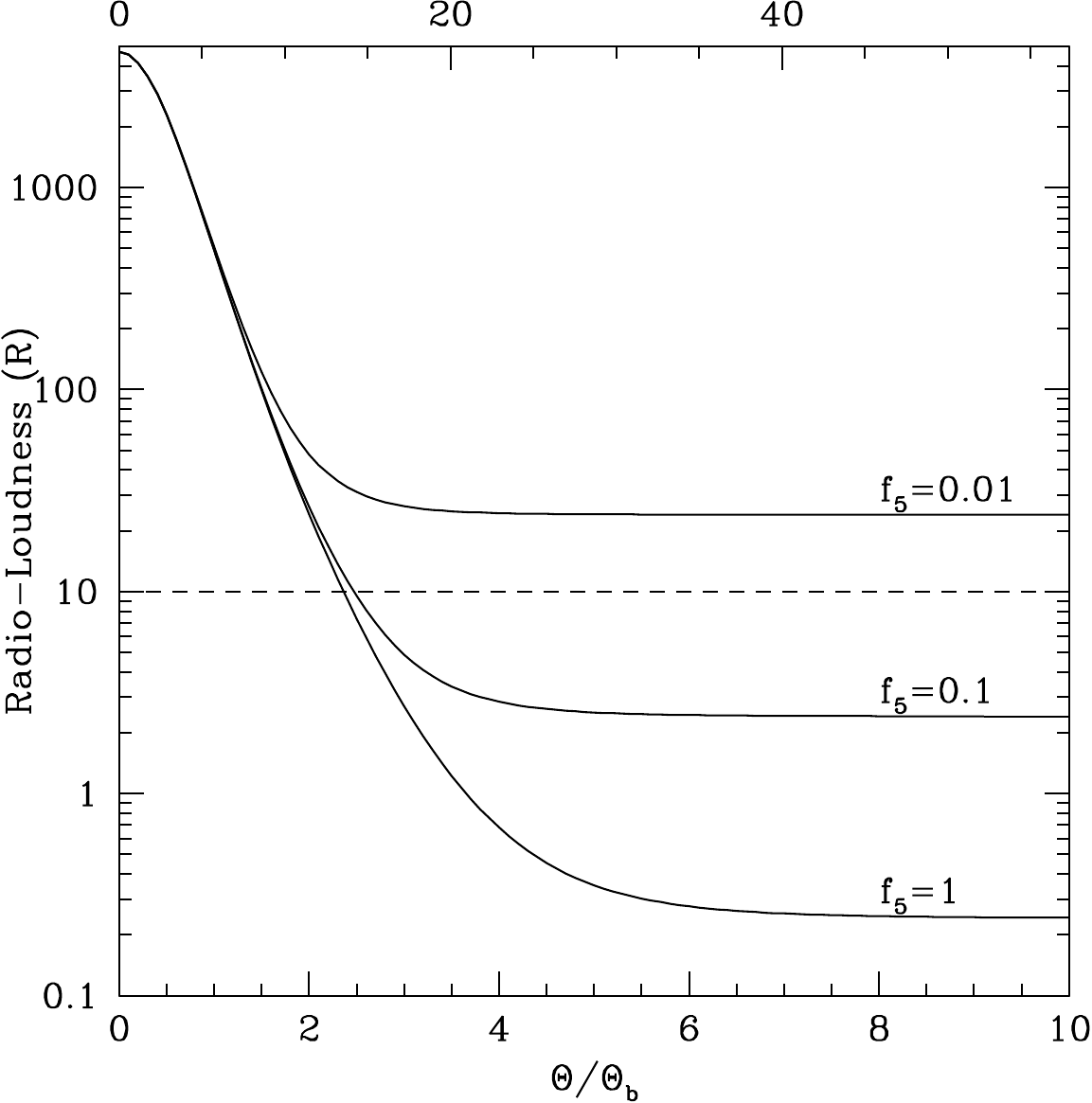}
    \caption{Radio-loudness vs the angle between the relativistic jet direction and the observer given in units of $\Theta_b=1/\Gamma$ (bottom axis) or in degrees (top axis), starting from a blazar observed at $\Theta$=0 deg with R=4700 and assuming $\Gamma$=10, p=3.3 and different values of the fraction between the intrinsic jet luminosity and the extended luminosity computed at 5~GHz (f$_5$).
    }
              \label{R_theta}
    \end{figure}

\subsection{The role of orientation}
If the high-z QSOs with R$<$10 are jetted sources, we need an explanation for their low radio emission, compared to radio-loud objects.  Since the radio emission (or at least part of it) in jetted QSOs is highly anisotropic, due to the beaming, while the optical/UV is usually dominated by the emission from the accretion disk, it is reasonable to consider the possibility that the high-z QSOs with low values of R are simply misaligned sources, where the radio emission is relatively weak because it is less-beamed (or even de-beamed) compared to oriented sources (blazars). 
The relevance of orientation on the observed value of radio-loudness has been widely discussed in the past years (e.g. see \citealt{Urry1995}). The general conclusion is that its effect is not important when the radio emission is dominated by the extended, non-beamed components, i.e. in lobe-dominated radio-galaxies, but it is expected to be significant for relatively small viewing angles, when the core becomes more relevant. 

In order to show this effect, we consider a plasma moving within a jet with a bulk velocity $v$ corresponding to a Lorentz factor $\Gamma$. 
The emission (P$^{jet}$) at a given frequency (e.g. 5~GHz) observed at a given angle ($\theta$) from the jet direction, is relativistically beamed according to the following equation (e.g. \citealt{Urry1995}):
\begin{equation}
    P^{jet}_{5GHz}(\theta)=\mathcal{P}^{jet}_{5GHz}\delta^p(\theta)
\end{equation}
where $\mathcal{P}^{jet}_{5GHz}$ is the jet radio power at 5~GHz in the source's rest frame, $\delta(\theta)$ is the Doppler factor\footnote{the Doppler factor is defined as $\delta=[\Gamma(1-\beta cos\theta)]^{-1}$, where 
$\Gamma$ and $\beta$ are the Lorentz factor and the ratio between the bulk velocity and the 
light speed, respectively.} and p is a parameter defining the jet properties. For instance, if $\alpha$ is the radio spectral index of the emitting source, we expect p=3+$\alpha$ for a moving isotropic source and p=2+$\alpha$ for a continuous jet (see Appendix B of \citealt{Urry1995} for more details). In the following, we assume the value of p estimated in \citet{Caccianiga2024} for high-z QSOs (p=3.3) and $\Gamma$=10.

We also assume that there is an extended, non-beamed radio emission, and that the ratio between the jet radio power at 5~GHz in the source's rest frame and the extended radio power at the same frequency ($\mathcal{P}^{ext}_{5GHz}$) is given by the parameter $f_5$ defined as:
\begin{equation}
    f_5 = \frac{\mathcal{P}^{jet}_{5GHz}}{\mathcal{P}^{ext}_{5GHz}}
\end{equation}

We further assume that the optical magnitude is not affected by beaming, i.e. that the optical/UV emission is mainly produced by the accretion disk with little or no contribution from the jet. Under these conditions, both the observed total radio power ($P^{tot}_{5GHz}(\theta)=P^{jet}_{5GHz}(\theta)+P^{ext}_{5GHz}$) and the value of R will decrease as we increase the viewing angle, following the beaming formula:
\begin{equation}
    P^{tot}_{5GHz}(\theta)=P^{tot}_{5GHz}(0)\Bigg(\frac{\delta^p(\theta)+1/f_5}{\delta^p(0)+1/f_5}\Bigg)
\end{equation}
and:
\begin{equation}
    R(\theta)=R(0) \Bigg(\frac{\delta^p(\theta)+1/f_5}{\delta^p(0)+1/f_5}\Bigg)
\end{equation}
where $P^{tot}_{5GHz}(0)$ and $R(0)$ are, respectively, the total radio power and the radio-loudness of a source observed exactly along the jet direction ($\theta$=0~deg).
In Fig.\ref{R_theta} we consider, as starting point, the value of radio-loudness (4700) of the most powerful object in the sample, HZQJ213412.0-041909 (alias PMN J2134-0419 at z=4.35) that has been classified as blazar (\citealt{ Sbarrato2014a, Cao2017}). Among the blazars included in the sample this is likely the one observed at the smallest angle and we can assume $\theta\sim$0 deg.

From figure~\ref{R_theta} we see that orientation affects the observed value of R only for relatively small angles ($\theta<$4$\Theta_b$ which corresponds to $\sim$20 deg for $\Gamma$=10). Moreover, depending on the value of f$_5$, the observed radio-loudness can reach low values at high viewing angles: while in sources with f$_5\sim$0.01, the value of R remains always well above the radio-loud/radio-quiet threshold, even for large viewing angles, in sources with f$_5\sim$0.1-1 the radio-loudness falls below 10 already for angles of $\sim$15-17 deg.

The value of f$_5$ estimated for powerful, low redshift, FR~II is of the order of 0.01 (\citealt{Urry1995, Giovannini2001}) but less powerful objects show higher values, reaching 0.1 or more (\citealt{Giovannini2001}). Several classes of radio-galaxies, like gigahertz peaked-spectrum (GPS) and compact Steep-spectrum (CSS) sources (\citealt{ODea1998,  Hardcastle2020, ODea2021}) or the low-luminosity FR0s (\citealt{Baldi2023}), are intrinsically compact and show weak extended radio emissions.
These objects are expected to have the highest values of f$_5$ 
Although we do not have a direct measurement of f$_5$ in these type of radio-galaxies, the statistical analysis of a sample of CSSs, discussed by \citet{Berton2016e}, suggests values close to $\sim$1.

At high redshift, many radio-emitting QSOs show properties similar to GPS radio-sources, like intrinsically compact radio morphologies (\citealt{Frey2008, Frey2010, Frey2011, Frey2018, Gabanyi2021, Zhang2021}) and/or peaked radio-spectra (\citealt{Coppejans2016a, Coppejans2017, Belladitta2023a}).
In addition, the extended emission at high redshift could be significantly damped due to the interaction between the relativistic electrons in the most extended parts of the jet and the photons from the Cosmic Microwave Background (CMB), whose density increases as (1+z)$^4$ (e.g. \citealt{Ghisellini2009b, Paliya2020b, Ighina2021c}). 
It is therefore reasonable to expect that in high-redshift jetted QSOs the value of f$_5$ is much higher compared to local extended radio-galaxies, and likely close to 0.1-1. If this is the case, Fig.~\ref{R_theta} shows that high-z jetted QSOs can have very low values of radio-loudness, well within the typical range observed in radio-quiet sources. 

We now want to test if orientation can explain the actual range of radio-loudness and radio power observed in the high-z QSOs detected in EMU and shown in Fig.~\ref{lum_R}. Instead of using only one blazar, as done for Fig.~\ref{R_theta}, we consider all the 10 confirmed blazars currently found in the sample to have a more stable starting point. These blazars are likely observed at different angles, between 0~deg and 1/$\Gamma\sim5.7$~deg with respect to the jet direction. As explained in \citet{Caccianiga2024}, the average Doppler factor of a sample of blazar is $\delta\sim$1.385$\Gamma$ that corresponds to an average viewing angle of $\sim$3.9~deg for $\Gamma$=10. Considering this angle and using the average values of P$^{tot}_{5GHz}$ and R computed for all the 10 blazars (open star on Fig.~\ref{lum_R}) as starting point, we calculated the expected values of radio power and R at different viewing angles from $\theta$=0 deg up to 4$\theta_b$, assuming f$_5$=1 (magenta line on Fig.~\ref{lum_R}). We see that the range of values of P$^{tot}_{5GHz}$ and R observed in the sample can be easily produced by a variation of viewing angle from 0 up to $\sim$3.5$\theta_b\sim$20~deg. The large spread around this line is mainly due to the wide range of intrinsic (in the source's rest-frame) jet/disk power ratios that likely characterize these sources. 
If we assume f$_5$=0.1, we still recover most of the observed range of power/radio-loudness but using larger angles ($\sim$30-40~deg). On the contrary, with f$_5$=0.01 it would not be possible to explain the sources with R$<$10 just by increasing the viewing angle since the presence of a strong extended emission keeps the values of R above 10.
This means that, for values of f$_5$ expected at high redshift (f$_5\sim$0.1-1), the objects with the lowest values of P$^{tot}_{5GHz}$ and R could be naturally explained by an orientation effect. Fig.~\ref{lum_R} also suggests that, depending on the actual value (or values) of f$_5$, there might be more jetted QSOs at even fainter powers (and lower radio-loudness) that are not yet detectable at the sensitivity of EMU. Some examples of high-z QSOs, formally classified as radio-quiet, but showing evidences for the presence of relativistic jets are now emerging from deep radio follow-ups (e.g. \citealt{Sbarrato2021a, Sbarrato2022}).

Currently, we do not have radio data with a resolution high enough to directly test if the sources in the sample are really compact radio sources with a relatively weak extended emission. The EMU resolution ($\sim$15$\arcsec$) corresponds to $\sim$80-100 kpc at z=4-6.6. Most of the sources are not spatially resolved or marginally extended with no evidence for large double morphologies. However, we need sensitive VLBI observations with resolutions of a few~m.a.s. in order to assess the actual sizes of these sources and to characterize their morphologies. This type of observation will be soon possible thanks to SKA-VLBI  (see Spingola et al. 2026 in prep).  

Similarly, the analysis of the radio spectrum of the sources, in particular of those with R$<$10, would be instrumental to assess their nature. 
To date, non-simultaneous radio data at a frequency different from 0.943~GHZ are available only for 12 QSOs with R$<$10 (i.e. 17\% of the radio-quiet QSOs in the sample). 
These limited and non-simultaneous data preclude a definitive assessment of the spectral nature of the $R<10$ sources. Dedicated multi-frequency radio follow-ups will be necessary to obtain a full characterization of the radio spectral properties of the radio-quiet QSOs in the sample.

\subsection{The predicted number of misaligned jetted QSOs}
\label{sec:pred}
\begin{figure}
   \centering
    \includegraphics[width=9cm, angle=0]{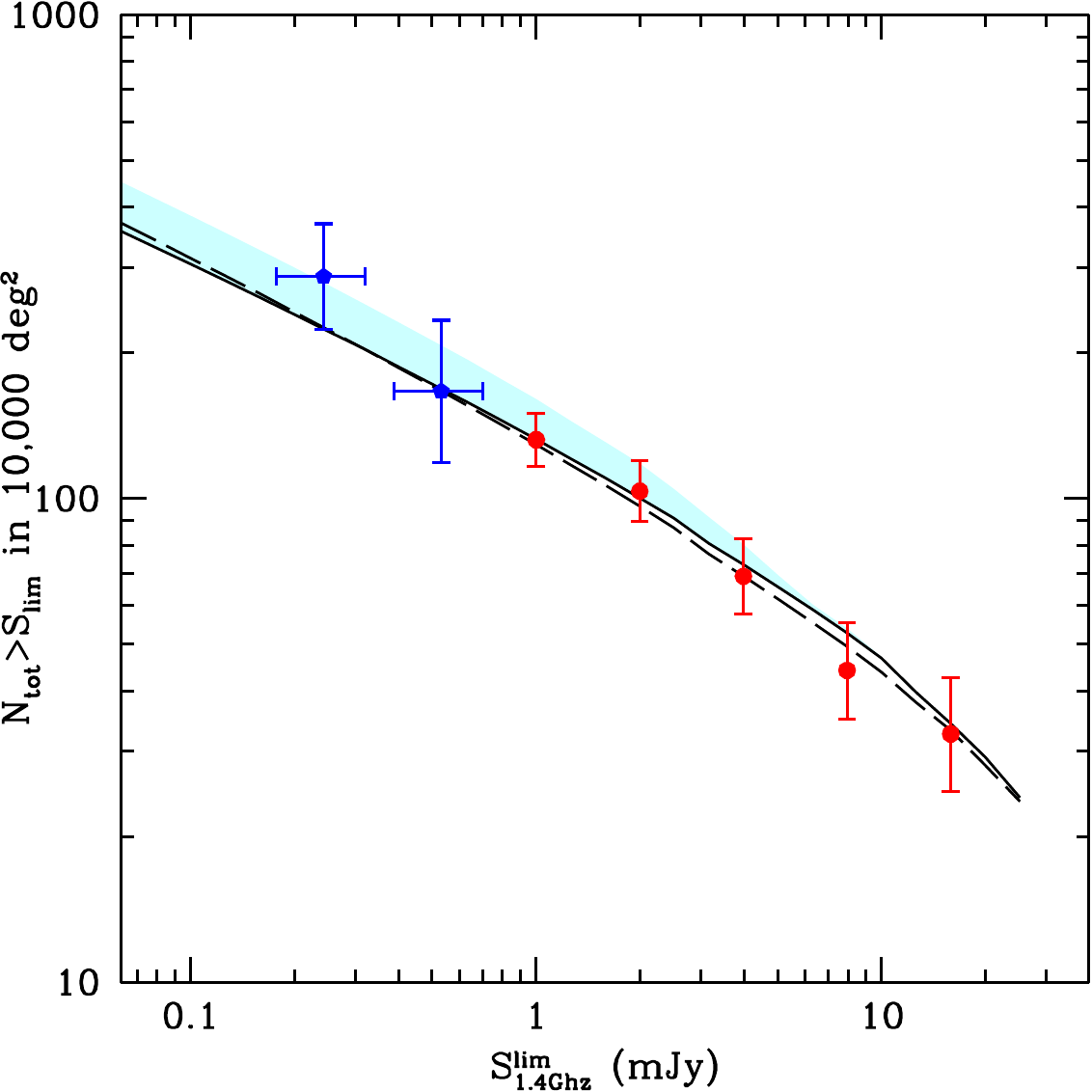}
    \caption{Observed number, re-scaled to a sky area of 10 000 sq. degrees, of z$\geq$4 jetted type1 QSOs (blazars+misaligned sources) with mag$_{drop}\leq$21 computed at different radio flux density limits, using the FIRST sample (\citealt{Caccianiga2024}, red points) and the EMU sample in the SDSS area (blue points). The continuous line represents the predictions 
    based on the high-z blazars currently detected at 1.4 GHz in the FIRST sample (mostly at flux densities above 10 mJy) described in \citet{Caccianiga2024}. The light-blue shaded area indicates uncertainties on these predictions. 
    The vertical error-bars are based on the Poissonian uncertainties, while the horizontal error-bars show the uncertainty on the radio density limit translated from the EMU frequency (0.943~GHz) to 1.4~GHz. Adapted from \citet{Caccianiga2024}.
    }         \label{fig_first_with_EMU}
    \end{figure}

The discussion presented in the previous section, based on Fig.~\ref{lum_R}, and the considerations on the total radio powers, suggests that the high-z QSOs detected in EMU could be jetted QSOs, even if the observed values of radio-loudness are relatively low. 
At present, we do not have enough data to directly test this hypothesis for each object in the sample. We can, however, quantitatively test the general consistency of this scenario in terms of space density: according to the beaming model (\citealt{Urry1995}) we expect that, in a flux limited sample, for a given number of oriented sources (blazars) there must be a well-defined number of misaligned ones. 
 In \citet{Caccianiga2024} we presented a method, originally proposed by \citet{Ghisellini2016}, to predict the expected number of misaligned sources in a flux limited sample on the basis of the observed flux densities of blazars. In particular, we used a well-defined sample of 20 z$>$4 blazars with mag$_{drop}\leq$21 selected from the combination of the FIRST survey (\citealt{Becker1995}) and the SDSS to predict the expected number of jetted QSOs (both oriented and misaligned) at different flux density limits. These predictions are shown in Fig.~\ref{fig_first_with_EMU}.

In order to compare these predictions with the number of sources found in the EMU sample, we have restricted the analysis to the sky area defined as 9h$\leq$RA$\leq$16h and $-$2.5deg$\leq$DEC$\leq$60deg (see Fig.~\ref{sky_dist}) which is fully covered by the SDSS spectroscopic observations. In this area, we are confident that most of z$\geq$4 QSOs with mag$\leq$21 have been discovered (\citealt{Schindler2017a, Caccianiga2024}). As for the radio flux density limit, the EMU catalogue contains sources as faint as 0.1~mJy/beam or even lower but the completeness at these fluxes is low. 
As explained in \citet{Hopkins2025}, the northern tiles, like the ones falling in the SDSS area, are observed for shorter time (5h) compared to the southern ones and the rms is therefore higher, from a median rms of 0.030 mJy/beam to 0.046 mJy/beam (\citealt{Hopkins2025}).
We have verified that in 90\% of cases the value of rms reported in the EMU catalogue for the images within the SDSS area is better than 0.064 mJy/beam. For this reason, we consider here only the sources with S$_{0.943GHz}>$5$\times$0.064=0.32~mJy/beam where the sky coverage can be considered reasonably flat. 
In total, we count 19 z$\geq$4 QSOs, with mag$_{drop}\leq$21 and S$_{0.943GHz}\geq$0.32~mJy/beam (corresponding to S$_{1.4GHz}$=0.24 mJy/beam assuming $\alpha_R$=0.7), detected  in  $\sim$660~deg$^2$ i.e. 0.029 objects/deg$^2$ i.e. 290 objects in 10,000 deg$^2$.  Considering larger flux densities (S$_{0.943GHz}\geq$0.7) we obtain $\sim$170 objects per 10,000 deg$^2$. Both values are consistent with the predictions (blue points in Fig.~\ref{fig_first_with_EMU}).

This good consistency further supports the scenario in which the high-z QSOs currently selected in EMU are jetted sources, mostly observed outside the ``beaming cone" (1/$\Gamma$), probably at 10-20 degrees from the jet direction. They are the misaligned counterparts of the powerful blazars observed at much higher flux densities ($>$10~mJy), like the ones selected in the FIRST survey (\citealt{Caccianiga2024}).
Conversely, if the high-redshift QSOs with $R<10$ detected by EMU belonged to a completely different population of sources, not powered by relativistic jets, we would have to explain why we do not observe the misaligned QSOs predicted by the beaming model.

Once completed, the EMU survey is expected to contain hundreds of high-z QSOs with mag$_{drop}<$21. This final sample will permit a more accurate comparison between the observed number of sources and the predictions based on the beaming model. 

\section{Summary and conclusions}
We have presented a sizeable sample of 808 QSOs with 4$\leq$z$\leq$9, assembled from the literature and from the SDSS, falling in the sky area covered by the EMU survey as of the end of February 2026 (about 9000 deg$^2$). This is one of the largest samples of z$\geq$4 radio-emitting QSOs for which deep radio data ($\sim$ mJy/beam at 0.943~GHz) are available. About 15\% of these objects are detected ($5\times \mathrm{rms}$) by EMU. These objects were previously known from the literature as high-z QSOs but, in the majority of cases, not as radio-emitting sources. Among these 124 QSOs, there are 15 objects with 5.6$\leq$z$\leq$6.6, nine of which are detected for the first time in the radio.
In addition, we tentatively detected at lower significance (between $5 \times \mathrm{rms}$ and $5 \times \mathrm{rms}$) four more z$>$5.5 QSOs.

The main findings of the sample analysis can be summarized as follows:

\begin{itemize} 

\item the fraction of radio-loud QSOs (R$>$10) is constant from z=4 up to z=6 and equal to 8$\pm$1\%,  thus confirming what has been previously found in the literature;

\item The fraction of radio-detected QSOs at high-z is significantly higher than 8\% and strongly dependent on the optical magnitude, reaching 35\% for the brightest objects in the sample (mag$_{drop}\sim$18).

\item On the basis of the high values of radio power observed in the high-z QSOs detected by EMU (P$_{5GHz}$=5$\times$10$^{24}$-5$\times$10$^{27}$ W Hz$^{-1}$), we have argued that also the QSOs with R$<$10, or most of them, are fully consistent with hosting a relativistic jet, probably slightly misaligned with respect to the line-of-sight i.e. observed outside the ``beaming cone" ($\sim$1/$\Gamma$). We have also shown that this hypothesis is consistent with the predictions of the beaming model, based on the number of blazars currently detected at higher fluxes;

\item In order to explain the low values of R observed in many of the high-z QSOs detected by EMU we have suggested that the extended, non-beamed radio-emission in these QSOs could be, on average, significantly lower compared to what is observed at lower redshift in powerful radio-loud AGN. In particular, the jet-to-extended emission (in the source's rest frame) in these sources could be an order of magnitude, or more, lower than the value measured in local giant radio-galaxies and could be similar to that observed in ``young" AGN, like GPS/CSS sources, where the development of the radio lobes is still on-going. The interaction of the jet with the photons from the CMB at high redshift is also expected to contribute to the reduced luminosity of the extended components. 

\end{itemize}

\noindent
If many high-z QSOs hosting powerful jets are really hiding among sources formally classified as radio-quiet, then the comparison with low-z sample should be significantly revised. For instance, the significant decrease at high-z of the radio-loud fraction observed by \citet{Jiang2007} would not necessarily imply a decrease of the fraction of jetted QSOs. Rather, it could only reflect the fact that the extended radio emission becomes progressively less important at high redshift, making many misaligned QSOs appearing as radio-quiet. Similar conclusions about the inadequacy of the radio-loudness parameter to correctly identify jetted QSOs at high redshift have been discussed in \citet{Sbarrato2022}, based on the analysis of high-z blazars. 

Dedicated follow-ups of the high-z QSOs detected by EMU are now needed to confirm the results discussed here. In particular, sensitive VLBI observations will be instrumental to confirm the nature of these sources and to directly test the hypothesis that most of these high-z QSOs are characterized by a weak extended emission. In addition, simultaneous multi-frequency radio data will help to fully characterize their radio spectrum which is another important proxy for the nature of these sources. 
In the near future, deep sub-arcsecond observations with SKAO of the high-z AGN that are emerging from JWST observations will also be fundamental to establishing if the same high radio detection fraction is observed in optically fainter samples, likely hosting smaller SMBHs. 

\section*{Data availability}
The full Table~1, containing the entire list of z$>$4 QSOs detected in EMU is available in electronic form at the CDS.

Appendix A is available on Zenodo at https://doi.org/10.5281/zenodo.24462937

\begin{acknowledgements}
We thank the referee for useful comments and suggestions that helped improve the manuscript. 
We acknowledge financial contribution from the agreement ASI-INAF n. I/037/12/0 and n. 2017-14-H.0 and from INAF under PRIN SKA/CTA FORECaST. We acknowledge financial support from INAF under the project `QSO jets in the early Universe', Ricerca Fondamentale 2022 and under the project `Testing the obscuration in the Early Universe', Ricerca Fondamentale 2023. CS acknowledges financial support from INAF (Grants 1.05.12.04.04). 
J.A. and R.C acknowledge financial support from the Science and Technology Foundation (FCT, Portugal) through research grant UID/04434/2025 (DOI: 10.54499/UID/04434/2025).
This work has been partially supported by the ASI-INAF program I/004/11/4.
This scientific work uses data obtained from Inyarrimanha Ilgari Bundara, the CSIRO Murchison Radio-astronomy Observatory. We acknowledge the Wajarri Yamaji People as the Traditional Owners and native title holders of the Observatory site. CSIRO’s ASKAP radio telescope is part of the Australia Telescope National Facility (https://ror.org/05qajvd42). Operation of ASKAP is funded by the Australian Government with support from the National Collaborative Research Infrastructure Strategy. ASKAP uses the resources of the Pawsey Supercomputing Research Centre. Establishment of ASKAP, Inyarrimanha Ilgari Bundara, the CSIRO Murchison Radio-astronomy Observatory and the Pawsey Supercomputing Research Centre are initiatives of the Australian Government, with support from the Government of Western Australia and the Science and Industry Endowment Fund
This work is based on SDSS data.
Funding for the Sloan Digital Sky Survey V has been provided by the Alfred P. Sloan Foundation, the Heising-Simons Foundation, the National Science Foundation, and the Participating Institutions. SDSS acknowledges support and resources from the Center for High-Performance Computing at the University of Utah. The SDSS web site is \url{www.sdss.org}.
\end{acknowledgements}

\bibliographystyle{aa}
\bibliography{blazar_high_z}

\end{document}